\documentclass[a4paper,preprintnumbers,showpacs,twocolumn,superscriptaddress,nofootinbib,amsmath,amssymb]{revtex4-1}
\usepackage[dvips]{graphics}
\usepackage[hypertex]{hyperref}
\usepackage{color,hyperref}
\usepackage{times}
\usepackage{mathrsfs}
\hypersetup{
  colorlinks=true,        
  linkcolor=blue,         
  citecolor=magenta,      
  filecolor=magenta,      
  urlcolor=blue            
}

\def\beq{\begin{equation}}
\def\eeq{\end{equation}}
\def\bear{\begin{eqnarray}}
\def\ear{\end{eqnarray}}
\def\nn{\nonumber}
\def\L{\mathscr{L}}

\usepackage{enumitem}
\usepackage{graphicx,subfigure}
\usepackage{dcolumn}
\usepackage{bm}
\usepackage{color}

\begin{document}

\title{Relaxations of perturbations of spacetimes in general relativity coupled to nonlinear electrodynamics}

\author{Bobir Toshmatov}
\email{bobir.toshmatov@fpf.slu.cz}
\affiliation{Department of Physics, Nazarbayev University, 53 Kabanbay Batyr, 010000 Astana, Kazakhstan}
\affiliation{Institute of Physics and Research Centre of Theoretical Physics and Astrophysics, Faculty of Philosophy \& Science, Silesian University in Opava, Bezru\v{c}ovo n\'{a}m\v{e}st\'{i} 13, CZ-74601 Opava, Czech Republic}
\affiliation{Ulugh Beg Astronomical Institute, Astronomicheskaya 33, Tashkent 100052, Uzbekistan}

\author{Zden\v{e}k Stuchl\'{i}k}
\email{zdenek.stuchlik@fpf.slu.cz}
\affiliation{Institute of Physics and Research Centre of Theoretical Physics and Astrophysics, Faculty of Philosophy \& Science, Silesian University in Opava, Bezru\v{c}ovo n\'{a}m\v{e}st\'{i} 13, CZ-74601 Opava, Czech Republic}

\author{Bobomurat Ahmedov}
\email{ahmedov@astrin.uz}
\affiliation{Ulugh Beg Astronomical Institute, Astronomicheskaya 33, Tashkent 100052, Uzbekistan}
\affiliation{National University of Uzbekistan, Tashkent 100174, Uzbekistan}

\author{Daniele Malafarina}
\email{daniele.malafarina@nu.edu.kz}
\affiliation{Department of Physics, Nazarbayev University, 53 Kabanbay Batyr, 010000 Astana, Kazakhstan}

\begin{abstract}

Three well known exact regular solutions of general relativity (GR) coupled to nonlinear electrodynamics (NED), namely the Maxwellian, Bardeen and Hayward regular spacetimes, which can describe either a regular black hole or a geometry without horizons, have been considered. Relaxation times for the scalar, electromagnetic (EM) and gravitational perturbations of black holes (BHs) and no-horizon spacetimes have been estimated in comparison with the ones of the Schwarzschild and Reissner-Nordstr\"{o}m (RN) spacetimes. It has been shown that the considered geometries in GR coupled to the NED have never vanishing circular photon orbits and on account of this fact these spacetimes always oscillate the EM perturbations with quasinormal frequencies (QNFs). Moreover we have shown that the EM perturbations in the eikonal regime can be a powerful tool to confirm that (i) the light rays do not follow null geodesics in the NED by the relaxation rates; (ii) if the underlying solution has a correct weak field limit to the Maxwell electrodynamics (LED) by the angular velocity of the circular photon orbit.

\end{abstract}

\maketitle

\section{Introduction}\label{sec-intr}

On September 14, 2015, the first ever detection of gravitational waves (GWs) from the coalescence of two stellar mass BHs by LIGO scientific collaboration lead to birth of an entirely new field of astronomy -- GW astronomy~\cite{GW150914}. Afterwards, LIGO and VIRGO scientific collaborations have announced the detection of several GWs from the merger of stellar mass BHs~\cite{GW151226,GW170104,GW170814,GW170608} and neutron stars (NSs)~\cite{GW170817}. Ground-based GW detectors such as LIGO and VIRGO have a few kilometer-long arms and can only observe the GW sources whose radiation is emitted at frequencies in the deca- and hecta-Hz band. Therefore, the ground-based GW detectors are sensitive to coalescence of NSs and stellar mass BHs. On the other hand, GWs at very low frequencies have wavelength larger than the Earth size as frequency of the GW is proportional to inverse of mass of the source. In this case these GWs cannot be detected by the ground-based detectors. To detect them, large enough antennas, so-called space-based antenna, away from Earth's surface are required. Space-based GW detectors, such as LISA (Laser Interferometer Space Antenna) can have million kilometers long arms and they are atmosphere, turbulence and seismic noise free and therefore, be sensitive in the milli-Hz band. The space-based GW detectors are expected to be sensitive to coalescences of supermassive BH-BH and supermassive BH-NS~\cite{Yunes:LRR}. The coalescence of the BHs (NSs) in binary occurs into three phases: inspiral, merger and ringdown -- each of which is calculated by the different methods. The inspiral represents the early evolution of the close binary system and since the binary components are far enough away each other, it can be treated by the post-Newtonian (PN) approximation by expanding expressions in powers of small relative velocity $v/c$. In the phase of merger, strong and highly dynamical gravitational fields develop, which can be treated only via numerical relativity simulations. Finally, in the last, merger phase the final object relaxes to its equilibrium state by radiating GWs whose frequencies are called quasi-normal (QN), since they are complex and subject to decay through the imaginary part. The merger is calculable via perturbation theory (semi-) analytically. In the BH perturbations theory one obtains the wave equation by introducing the linear small perturbation to a fixed BH background spacetimes and solving the Einstein equations in the linear order of perturbations. A perturbed BH in its queue goes through the following three stages: transient, QNM ringdown, and power law tail. Where the transient phase strongly depends on the initial perturbations, while the ringdown is independent of the initial perturbations and characterized by the QNMs which encode information about the BH~\cite{Kokkotas,Rezzolla:review,Berti:review,Konoplya:review}.

The above discussion regards only the gravitational perturbations. However, for scalar and EM perturbations, also through the typical standard analysis, similar wave equations can be obtained despite the different underlying physics. The scalar~\cite{Nomura:PRD:2005,Fernando:PRD:2012,TASA:PRD:2015,Yang:PRD:2016}, gravitational~\cite{ChaverraPRD:93,SarbachPRD:67} and EM~\cite{TSSA:PRD:2018,TSA:PRD:98b} perturbations of regular BHs have been studied. One of the important properties of the perturbations is the relaxation time which is defined by the inverse of imaginary part of the QNMs, $\tau=1/\omega_i$. In this paper we aim to study in the eikonal regime the relaxation times of the scalar, electromagnetic and gravitational perturbations of the Maxwellian, Bardeen and Hayward regular BHs in general relativity coupled to NED. In the eikonal regime scalar and gravitational perturbations behave similarly, following the null geodesics of spacetime~\cite{CardosoPRD:79}, however, the EM perturbations of spacetimes in the NED behave differently~\cite{TSSA:PRD:2018,TSA:PRD:98b}, due the fact that in the NED light rays do not follow null geodesics of the spacetime, instead they follow null geodesics of the optical metric~\cite{Novello:PRD:2000,Novello:PRD:2001,ObukhovPRD:66,deOlivieraCQG:26,StuchlikIJMPD:24,Schee:JCAP2015,Schee:CQG2016}.

The paper is organized as follows: In Sec.~\ref{sec2} we present the regular BH solutions in general relativity coupled to the NED and study the main properties of the spacetime. In Sec.~\ref{sec3} the scalar, EM and gravitational perturbations of regular spacetimes in general relativity coupled to the NED are described and in the eikonal regime their propagations, relaxation times are studied in comparison with the ones of the Schwarzschild and RN spacetimes. Finally, we summarize our results in Sec.~\ref{sec4}. In this paper we mainly use the geometric units $c=G=1$, and adopt $(-, +, +, +)$ convention for the signature of the metric.

\section{Background}\label{sec2}

The action of a system of general relativity coupled to nonlinear electrodynamics is given as

\begin{eqnarray}\label{action}
S = \frac{1}{16 \pi }\int d^4 x \sqrt{ -g }\left( R - \mathscr{L}(F) \right)\ ,
\end{eqnarray}
where $g$ is the determinant of the metric tensor, $R$ is the Ricci scalar, and $\L$ is the Lagrangian density describing the NED theory. $F\equiv F_{\alpha\beta}F^{\alpha\beta}$, with the EM field tensor being $F_{\alpha\beta}=\partial_\alpha A_\beta-\partial_\beta A_\alpha$, with $A^\alpha$ the 4-potential. Since $F_{\alpha\beta}$ is antisymmetric, it has only six nonzero components.

The covariant equations of motion are written in the form
\begin{eqnarray}
&&G_{\alpha\beta}=T_{\alpha\beta},  \label{eq-motion1}\\
&&\nabla_\beta\left(\L_F F^{\alpha\beta}\right)=0\ ,   \label{eq-motion2}
\end{eqnarray}
where $T_{\alpha\beta}$ and $G_{\alpha\beta}=R_{\alpha\beta}-Rg_{\alpha\beta}/2$ are the energy-momentum tensor of the NED field and the Einstein tensor, respectively. The energy-momentum tensor of the NED is determined by the relation
\begin{eqnarray}\label{em-tensor}
T_{\alpha\beta} = 2 \left(\L_F
F_\alpha^\gamma F_{\beta\gamma}
- \frac{1}{4}
g_{\alpha\beta} \L \right)\ ,
\end{eqnarray}
where $\L_F=\partial_F\L$.

Let us consider the line element of the static, spherically symmetric BH given in the form
\begin{eqnarray}
ds^2 = - f( r) d t^2 + \frac{ dr^2}{f (r)}
+r^2 \left( d \theta^2+ \sin^2 \theta  d\phi^2 \right)\ , \label{line-element}
\end{eqnarray}
where GR and NED evaluate the lapse function $f(r)$.

In general, the EM 4-potential can be written in the following form:
\begin{eqnarray}\label{ansatz}
\bar{A}_\alpha=\varphi(r)\delta_\alpha^t-Q_m\cos\theta\delta_\alpha^\phi\ ,
\end{eqnarray}
where $\varphi(r)$ and $Q_m$ are the electric potential and the total magnetic charge, respectively. Since the construction of the electrically and magnetically charged spacetime solutions have been shown in~\cite{Fan:PRD:2016,Bronnikov:PRD2017,TSA:PRDnew}, we here do not report derivation of the solution, instead we specify the model of NED and perform the further calculations. In the following we consider a generic class of magnetically charged regular BH solutions, which is given by the function~\cite{Fan:PRD:2016,TSA:PRDnew}
\bear\label{metric-function}
f(r)=1-\frac{2Mr^{\mu-1}}{(r^\nu+q^\nu)^{\frac{\mu}{\nu}}},
\ear
corresponding to the lagrangian density
\bear\label{lagrangian}
\L=\frac{4\mu}{\alpha}\frac{(\alpha
F)^{\frac{\nu+3}{4}}}{\left[1+ (\alpha
F)^{\frac{\nu}{4}}\right]^{1+\frac{\mu}{\nu}}}\ ,
\ear
where $q$ is the magnetic charge parameter. Here $\mu\geq 0$ and $\nu>0$ are dimensionless constants and the value of $\mu$ characterizes the strength of nonlinearity of the EM field. Notice that $\mu=0$ corresponds to the absence of NED which reduces the spacetime to the Schwarzschild solution. Also taking $\mu\geq3$ ensures the regularity of the spacetime everywhere~\cite{Fan:PRD:2016}. Finally, $M$ is the gravitational mass. In this framework several classes of well-known regular BH solutions can be obtained such as, $\nu=1$, $\nu=2$, and $\nu=3$ correspond to the Maxwellian solution that corresponds to the Maxwell field in weak field regime, Bardeen-like, and Hayward-like solutions, respectively. Hereafter, we perform calculations in these three types of regular spacetimes and compare their behaviour relative to each other and to the Schwarzschild BH.

The main properties of these spacetimes have been studied in~\cite{Fan:PRD:2016,Toshmatov:PRD:2017}, so here we shall mention the most crucial points, such as horizons of the spacetimes, since these are important for our further calculations. The coordinate singularity so-called event horizon of the spacetime is defined by the divergence of the spacetime metric through $g_{rr}$ component of the spacetime metric, which in our case by $f(r)=0$. When $q=0$ one recovers the Schwarzschild spacetime in Schwarzschild coordinates, with the coordinate singularity at radius $r=2M$ and the curvature singularity at $r=0$. In the NED solutions the presence of the charge parameter decreases the radius of event horizon and determines the existence of an inner horizon close to the center. With increasing value of the charge parameter the outer horizon's location decreases while the inner horizon's location increases. For a specific value of $q$ we obtain an extreme value for the horizon radius where the two horizons coincide. This corresponds to the solution of equations $f(r)=0$ and $f'(r)=0$. For values of the charge parameter above the extremal one, both horizons disappear and the spacetime no longer represents BH, instead it represents no-horizon spacetime. By solving $f=0=f'$ for $\mu=3$ we find two equations,
\bear
(r^\nu+q^\nu)^{3/\nu}-2Mr^2=0\ , \quad r^\nu-2q^\nu=0.
\ear
By solving them simultaneously we find the values $r_{ext}$ and $q_{ext}$ that denote the boundary of the bh case with the horizonless case. These values are
\begin{itemize}
\item Maxwellian BH: $(16/27\approx0.5926, 8/27\approx0.2963)$;
\item Bardeen BH: $(4\sqrt{2}/\sqrt{27}\approx1.0887, 4/\sqrt{27}\approx0.7698)$;
\item Hayward BH: $(4/3\approx1.3333, 4/(3\times2^{1/3})\approx1.0583)$,
\end{itemize}
\begin{figure}[h]
\centering
\includegraphics[width=0.48\textwidth]{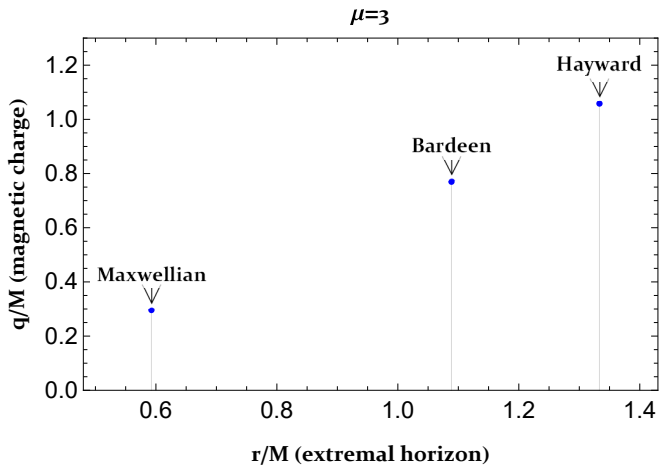}
\caption{\label{fig-extremal} Boundaries between the Maxwellian, Bardeen and Hayward regular BHs and the no-horizon spacetimes in the parametric space. The blue points correspond
to the extremal BHs ($r_{ext}/M$, $q_{ext}/M$). Where shaded regions represent the possible values of the magnetic charge parameter for the spacetime to represent BHs.}
\end{figure}
In Fig.~\ref{fig-extremal} the boundaries of BH and no-horizon Maxwellian, Bardeen and Hayward regular spacetimes are presented for the case of $\mu=3$~\footnote{Boundaries of the Maxwellian regular BHs and no-horizon spacetimes for different values of $\mu$ were studied in~\cite{TSA:PRD:98b}.}.

Thus, possible values of the charge parameter for the spacetimes~(\ref{metric-function}) with $\mu=3$ to represent the BHs lay in the following ranges: for Maxwellian BH $q/M\in[0,0.2963]$, for Bardeen BH $q/M\in[0,0.7698]$, and for Hayward BH $q/M\in[0,1.0583]$. Since these ranges are different, to facilitate the comparison, we normalize charge parameter as $Q_n\equiv q/q_{ext}$, and for the BH regime of the spacetime $Q_n$ lays in the range $Q_n\in[0,1]$. $Q_n\in[1,\infty)$ corresponds to the no-horizon spacetimes.

One of the astrophysically important orbits around BHs are light rings (photon spheres). It is well-known fact that in the LED and other NED nonrelated spacetimes light ray always follows the null geodesics and circular null geodesics (CNG) of the spacetime~(\ref{line-element}) is located at $r_c$ that is determined by solving equation ${\rm CNG}\equiv2f(r_c)-r_c f'(r_c)=0$. In our case it takes the form
\bear\label{null-geod}
M r_c^{\mu -1} \left(q_c^{\nu}+r_c^{\nu }\right)^{-\frac{\mu}{\nu}-1} \left[(\mu-3)q_c^{\nu }-3 r_c^{\nu}\right]+1=0\ ,
\ear
The radius $r_c$ can be considered as the one of the circular massless neutrino orbit~\cite{SS:EPJC:2018}. As in the case of the event horizon, presence of the charge parameter evaluates the inner and outer CNG orbits in regular spacetimes, with increasing the value of charge parameter they approach each other and before disappearing, at the extremal CNG which is solution of the equation ${\rm CNG}=0$ and ${\rm CNG},_{r}=0$, they coincide. The extremal CNG is located at the following coordinates of parametric space ($r_{c/ext}/M,q_{c/ext}/M (Q_n)$):
\begin{itemize}
\item Maxwellian spacetime: $(0.9492, 0.3164 (1.0679))$;
\item Bardeen spacetime: $(1.7173, 0.8586 (1.1154))$;
\item Hayward spacetime: $(2.0833, 1.2183 (1.1513))$.
\end{itemize}
and they follow the relation
\bear\label{extr-cng}
&&r_{c/ext}=\\&&\left(\frac{\mu(\nu+4)+\sqrt{\mu}\sqrt{\mu(\nu+2)^2+4\nu(\mu-3) }-6}{6}\right)^{1/\nu}q_{c/ext},\nonumber
\ear
Since we are mainly focusing our attention on the minimal value $\mu=3$ that makes the spacetimes regular then, expression~(\ref{extr-cng}) takes more compact form as
\bear\label{extr-cng1}
r_{c/ext}=\left(\nu+2\right)^{1/\nu}q_{c/ext}.
\ear

\begin{figure*}[th]
\centering
\includegraphics[width=0.48\textwidth]{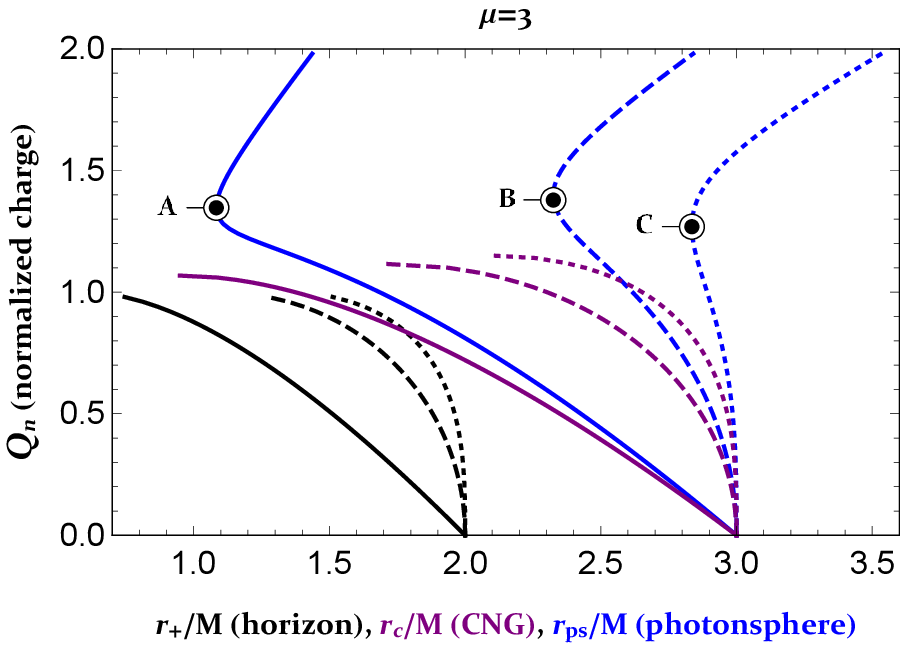}
\includegraphics[width=0.48\textwidth]{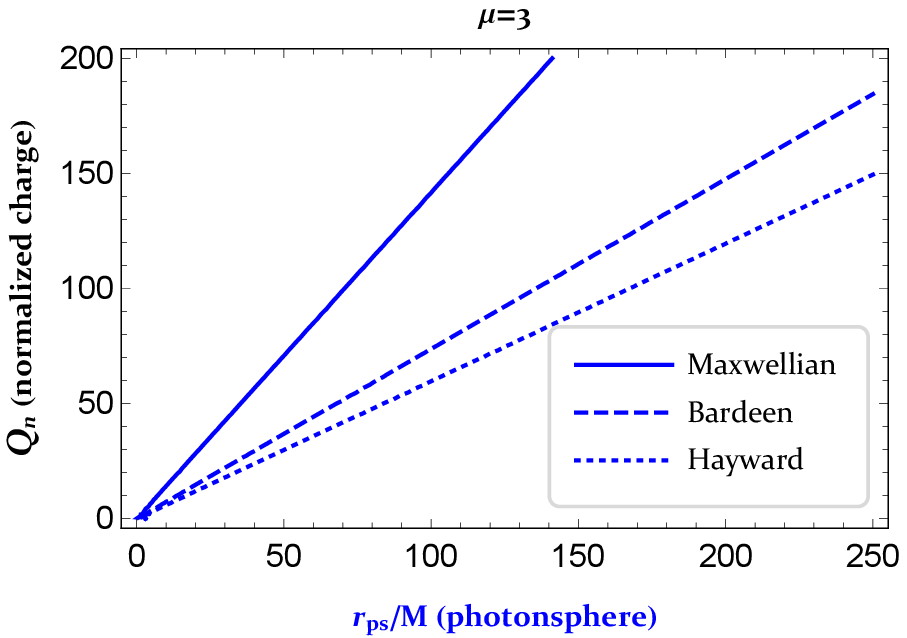}
\caption{\label{fig-rps} Left panel: dependence of radii of event horizon (black), circular null geodesics (purple), and light ring (blue) of the Maxwellian ($\nu=1$, solid), Bardeen ($\nu=2$, dashed), and Hayward ($\nu=3$, dotted) regular BHs on the normalized charge parameters. Where points $A$, $B$ and $C$ correspond to minimal radii of photonspheres in corresponding spacetimes. Right panel: radii of the light rings in the Maxwellian, Bardeen, and Hayward spacetimes for the large value of normalized charge parameter.}
\end{figure*}

By comparing the above given values with the ones of the extremal horizons, or seeing Fig.~\ref{fig-rps}, one can make sure that even no-horizon spacetime can possess the circular null geodesics, for a limited range of values of the charge parameters as
\begin{itemize}

\item Maxwellian no-horizon spacetime: $Q_n\in(1, 1.0679]$ (or $q/M\in(0.2963, 0.3164]$) at $r_c/M\in[0.9492, 1.3731)$;

\item Bardeen no-horizon spacetime: $Q_n\in(1, 1.1154]$ (or $q/M\in(0.7698, 0.8586]$) at $r_c/M\in[1.7173, 2.3012)$;

\item Hayward no-horizon spacetime: $Q_n\in(1, 1.1513]$ (or $q/M\in(1.0583, 1.2183]$) at $r_c/M\in[2.0833, 2.6524)$;

\item RN naked singularity spacetime: $Q_n\in(1, 1.0607]$ (or $q/M\in(1, 1.0607]$) at $r_c/M\in[1.5, 2)$.

\end{itemize}

However, in the NED light rays do not follow the
null geodesics of the original metric, instead, they propagate along the null geodesics of the effective (or optical)
metric which is given by~\cite{Bronnikov:PRD:2001,Novello:PRD:2000,Novello:PRD:2001}
\bear\label{eff-metric}
ds^2 =-\frac{1}{\L_F}\left[f(r)dt^2-\frac{dr^2}{f(r)}\right]
+\frac{r^2}{\Phi}d\Omega^2\ ,
\ear
where $\Phi=\L_F+2F\L_{FF}$. Thus, the photon sphere of spacetime~(\ref{line-element}) is located at the unstable circular null geodesics of the metric~(\ref{eff-metric}) that is determined by solving equation
\bear\label{photon-orbit}
\left(\frac{r^2}{\Phi}\right)_{ps}'\frac{f_{ps}}{\L_{Fps}}- \frac{r_{ps}^2}{\Phi_{ps}}\left(\frac{f}{\L_F}\right)_{ps}'=0\ .
\ear
Let us write Eq.~(\ref{photon-orbit}) for considered spacetimes. For the Maxwellian spacetime with $\mu=3$, $\nu=1$
\bear\label{ps-maxwellian}
12M-\frac{28Mq}{r}+\frac{\left(6 q^2+7 q r-4 r^2\right) (q+r)^3}{r^4}=0.
\ear
For the Bardeen spacetime $\mu=3$, $\nu=2$
\bear\label{ps-bardeen}
18M-\frac{52 q^2}{r^2}+\sqrt{\frac{q^2}{r^2}+1} \left(\frac{8 q^6}{r^5}+\frac{24 q^4}{r^3}+\frac{10 q^2}{r}-6 r\right)=0.\nonumber\\
\ear
For the Hayward spacetime $\mu=3$, $\nu=3$
\bear\label{ps-hayward}
24M-8r+\frac{10 q^9}{r^8}+\frac{39 q^6}{r^5}+\frac{21 q^3 (r-4M)}{r^3}=0.
\ear
Despite, Eqs.~(\ref{ps-maxwellian}),~(\ref{ps-bardeen}),~(\ref{ps-hayward}) are analytically not solvable, it is not difficult to check that for all values of $q$ they always have at least one real zero (see the right panel of Fig.~\ref{fig-rps}). In Fig.~\ref{fig-rps} the loci of the characteristic orbits are depicted. One can see that unlike the case standard LED or other NED not related spacetimes, in the considered regular spacetimes given by the line element (\ref{line-element}) with metric function (\ref{metric-function}) and $\mu=3$, $\nu=1,2,3$ the GR coupled to the NED theory (8) always gives nonvanishing circular photon orbit. It must be noted that different radial coordinates denote different radial distances in different space-times. Therefore, in principle we cannot directly compare the values of CNGs for different metrics. In the cases under consideration here, radial distances are determined by the line elements (\ref{line-element}) and (\ref{eff-metric}), which depend upon the parameters $q$ and $\nu$ and reduce to known metrics in the limits of vanishing $q$. By evaluating the area of the surfaces of revolution for $t=cont.$ and $r=r_{0ps}$, it is easy to verify that for nonvanishing values of $q$ the radii of CNGs identify indeed spherical two-surfaces for every constant $t$ slice. Also, since such areas monotonically increase with $q$ we know that the behavior shown in Fig.~\ref{fig-rps} is qualitatively valid. The existence of circular photon orbits in all the Bardeen spacetimes has been firstly demonstrated in~\cite{SS:EPJC:2018}. Here we have shown that the Maxwellian, Bardeen and Hayward regular BH and no-horizon spacetimes in the NED model~(\ref{lagrangian}) have this property. One can see from Fig.~\ref{fig-rps} that increasing the value of the charge parameter the radius of the photon sphere decreases until the below listed points, then it starts to increase again. Therefore, these points correspond to minimal radii of photonspheres in corresponding spacetimes. Thus, the minimal photonspheres of the regular Maxwellian, Bardeen, and Hayward spacetimes are located at
\begin{itemize}
\item $A(r_{0ps}/M,Q_n)=(1.0834,1.3468)$;
\item $B(r_{0ps}/M,Q_n)=(2.3251,1.3986)$;
\item $C(r_{0ps}/M,Q_n)=(2.8355,1.2690)$,
\end{itemize}
and they correspond to the no-horizon spacetimes. Interestingly, we see that for a certain rage of radii, which in coordinate radii corresponds to $r \in (r_{0ps}, 3M]$ there may exist two photon spheres with the same radius and for different value of the charge parameter.

\section{Perturbations of spacetimes in general relativity coupled to NED}\label{sec3}

It is known that most of the problems concerning the perturbations of BHs can be reduced to a second order partial differential equation after decoupling of angular variables and considering the perturbations as harmonically time dependent, in the following form:
\bear\label{weq}
\left(\frac{\partial^2}{\partial x^2}+\omega_j^2-V_j(r)\right)\Psi_j(r)=0\ ,
\ear
where $j$ stands for $sc$ (scalar), $em$ (EM) and $gr$ (gravitational) perturbations, and $x$ is the tortoise coordinate that is defined as $dx=dr/f$. Let us present the explicit forms of potentials of scalar, $V_{sc}$, electromagnetic, $V_{em}$, and gravitational, $V_{gr}$, perturbations of the BHs in the NED  which are given separately with brief explanations in~\cite{TASA:PRD:2015,TSA:prep}.

\textit{\textbf{Scalar perturbations:}} Since the scattering potential of the test scalar field in the field of the spherically symmetric BHs is presented in~\cite{TASA:PRD:2015}, we will only provide the potential.
\bear\label{pot-scalar}
V_{sc}=f\left[\frac{\ell(\ell+1)}{r^2}+\frac{f'}{r}\right].
\ear
\textit{\textbf{Gravitational and EM perturbations:}} Both the axial and polar EM perturbations of the BHs in the NED that have been studied in our preceding papers~\cite{TSSA:PRD:2018,TSA:PRD:98b} are just special cases of the gravitational perturbation due to the fact that the EM one was neglected. Here we briefly give the general case where both perturbations are taken into account.

The EM perturbation of the magnetically charged (with the four-potential $\bar{A}_\phi=-Q_m\cos\theta$) BH in the NED is given as $A_\phi=\bar{A}_\phi+\delta A_\phi$ where
\bear\label{em-pert}
\delta A_\phi=\psi(r)e^{-i\sigma t}\sin\theta\partial_\theta P_k(\cos\theta)\ .
\ear
with $\sigma=\omega_{em}$ and $k$ is multipole number of the EM perturbations which is restricted by the condition $k\geq1$. The gravitational perturbation in the ``Regge-Wheeler" gauge is introduced as $g_{\mu\nu}=\bar{g}_{\mu\nu}+h_{\mu\nu}$ where
\begin{align}\label{gr-pert}
h_{\mu\nu}=&
\begin{pmatrix}
0 & 0 & 0 & h_0(r) \\
\ast & 0 & 0 & h_1(r) \\
\ast & \ast & 0 & 0 \\
\ast & \ast & \ast & 0
\end{pmatrix}e^{-i\omega t}\sin\theta\partial_\theta P_\ell(\cos\theta)\ ,
\end{align}
We insert the perturbed metric and EM 4-potential from
Eqs.~(\ref{em-pert})--(\ref{gr-pert}) into the Einstein and Maxwell equations in the equations of motion (\ref{eq-motion1}) and (\ref{eq-motion2}), and expand to first order in the perturbations. Thus, for the gravitational perturbations we obtain the following equations:
\bear
&&h_0''+i\omega h_1'+i\omega\frac{2h_1}{r} -\frac{h_0\left[\lambda+2 f+(r^2 f')'+\bar{\L} r^2\right]}{r^2f}=0,\label{gr-pert1}\nonumber\\
\\
&&i\omega h_0'-i\omega\frac{2h_0}{r}-\omega^2h_1+h_1 \frac{f\left[\lambda+(r^2 f')'+\bar{\L} r^2 \right]}{r^2}=0,\label{gr-pert2}\nonumber\\
\\
&&i\omega h_0=-f(h_1f)',\label{gr-pert3}
\ear
with
\bear
\lambda=(\ell+2)(\ell-1).
\ear
By eliminating $h_0$ from Eq.~(\ref{gr-pert2}) by using
the Eq.~(\ref{gr-pert3}), we arrive at the master equation~(\ref{weq}) for the gravitational perturbations with the potential
\bear\label{pot-grav}
V_{gr}=f\left[\frac{\ell(\ell+1)}{r^2}+\frac{r(rf')'+2(f-1)}{r^2}+\bar{\L}\right],
\ear
by introducing the following notation:
\bear
\Psi_{gr}=\frac{f}{r}h_1.
\ear

Let us now consider propagation of the EM perturbation~(\ref{em-pert}) in the perturbed spacetime~(\ref{gr-pert}). The equation that governs the EM perturbation~(\ref{eq-motion2}) with both perturbations~(\ref{em-pert}) and~(\ref{gr-pert}) appear to be independent of the gravitational perturbations in the linear order expansion. Then, the dynamics of the EM perturbation of the BHs in general relativity coupled to the NED without gravitational perturbations has been studied in our previous papers~\cite{TSSA:PRD:2018,TSA:PRD:98b}. Therefore, we report only the potential without giving the details of derivation as~\footnote{Since the EM perturbation is independent of the gravitational one, in the linear order expansion one can replace multipole number of the EM perturbation $k$ as $\ell$.}
\bear\label{pot-em}
&&V_{em}=f\times\\&&\left[\frac{\ell(\ell+1)}{r^2}\left(1+
\frac{4Q_m^2\bar{\L}_{\bar{F}\bar{F}}}{r^4\bar{\L}_{\bar{F}}}\right)
-\frac{f\bar{\L}_{\bar{F}}'^2-2\bar{\L}_{\bar{F}}
\left(f\bar{\L}_{\bar{F}}'\right)'}{4\bar{\L}_{\bar{F}}^2}\right], \nn
\ear
where
\bear
\Psi_{em}=\sqrt{\L_{F}}\psi_1.
\ear

In the large multipole numbers limit, the potentials~(\ref{pot-scalar}),~(\ref{pot-grav}), and~(\ref{pot-em}) take the following forms:
\bear
&&V_{gr}=V_{sc}=f\frac{\ell^2}{r^2}+O\left(\ell\right),\label{pot-grav1}\\
&&V_{em}=f\frac{\ell^2}{r^2}\left(1+
\frac{4Q_m^2\bar{\L}_{\bar{F}\bar{F}}}{r^4\bar{\L}_{\bar{F}}}\right)
+O\left(\ell\right).\label{pot-em1}
\ear
It is known that in the eikonal (large multipole number) regime the QNMs of all perturbations of any stationary, spherically symmetric and asymptotically flat black holes in any dimensions are characterized by the parameters of the circular null geodesics~\cite{CardosoPRD:79}, namely, the real part of the QNMs is determined by angular velocity of the unstable null geodesics, $\Omega_c$, while the imaginary part of the QNMs is determined by the instability
timescale of the orbit, the so-called Lyapunov exponent, $\lambda$, as
\bear\label{eikonal1}
\omega=\Omega_c\ell-i\left(n+\frac{1}{2}\right)|\lambda|,
\ear
\begin{figure*}[th]
\centering
\includegraphics[width=0.49\textwidth]{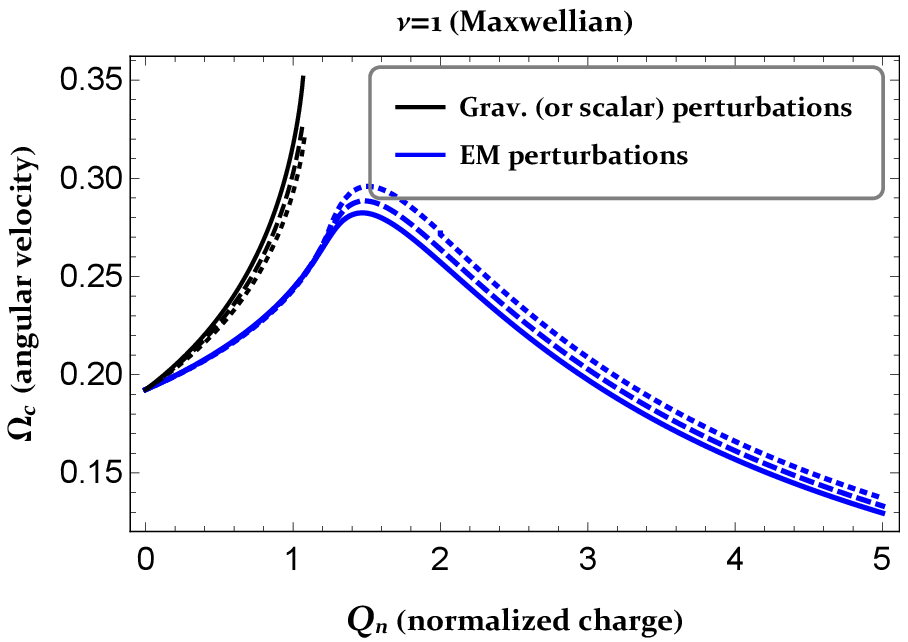}
\includegraphics[width=0.49\textwidth]{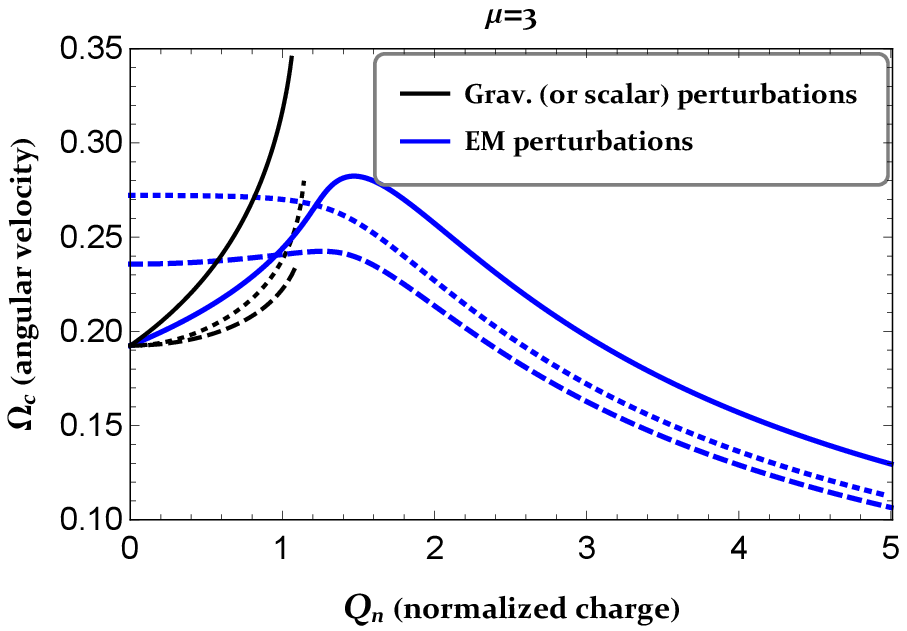}
\caption{\label{fig-real} Dependence of angular velocities of the circular null geodesics (black) and photon orbit (blue, thick) in the generic class of spacetimes~(\ref{line-element}) with metric function~(\ref{metric-function}) in general relativity coupled to the NED from normalized charge parameters. Left panel: the Maxwellian regular spacetimes ($\nu=1$) with different values of $\mu$ as $\mu=3$ -- solid, $\mu=5$ -- dashed, $\mu=12$ -- dotted curves. Right panel: the Maxwellian regular spacetimes ($\mu=3$, $\nu=1$) -- solid, the Bardeen regular spacetimes ($\mu=3$, $\nu=2$) -- dashed, the Hayward regular spacetimes ($\mu=3$, $\nu=3$) -- dotted curves.}
\end{figure*}
where $\Omega_c$ and $\lambda$ are determined by the spacetime metric~(\ref{line-element}) as
\bear
&&\Omega_c=\sqrt{\frac{f_c}{r_c^2}},\label{omega}\\
&&\lambda=\sqrt{-\frac{r_c^2}{2f_c}\left.\left(\frac{d^2}{dx^2}
\frac{f}{r^2}\right)\right|_{r=r_c}},\label{lambda}
\ear
where $r_c$ is the radius of the unstable null circular orbit which is determined by solving the equation $r_cf_c'-2f_c=0$. However, as it was mentioned in~\cite{TSSA:PRD:2018,TSA:PRD:98b}, the relation (\ref{eikonal1}) is not an universal feature of all stationary, spherically symmetric and asymptotically flat black holes in any dimensions, as it is not satisfied in several cases such as in the EM perturbations of BHs in NED~\cite{TSSA:PRD:2018,TSA:PRD:98b} and the gravitational perturbations of BHs in the Einstein -Lovelock theory~\cite{KonoplyaPLB:2017,KonoplyaJCAP:2017}.

\begin{figure*}[th]
\centering
\includegraphics[width=0.49\textwidth]{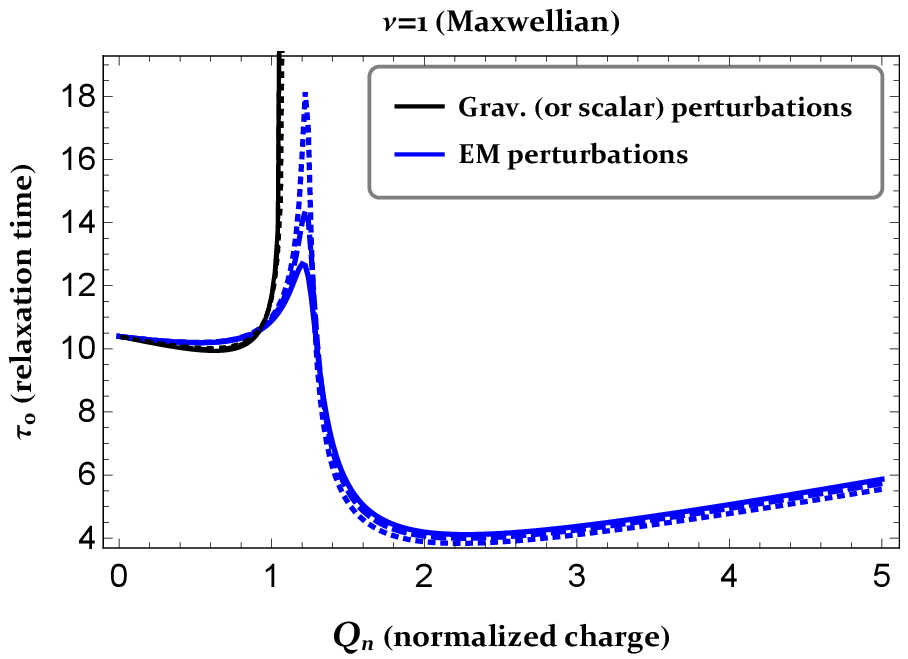}
\includegraphics[width=0.49\textwidth]{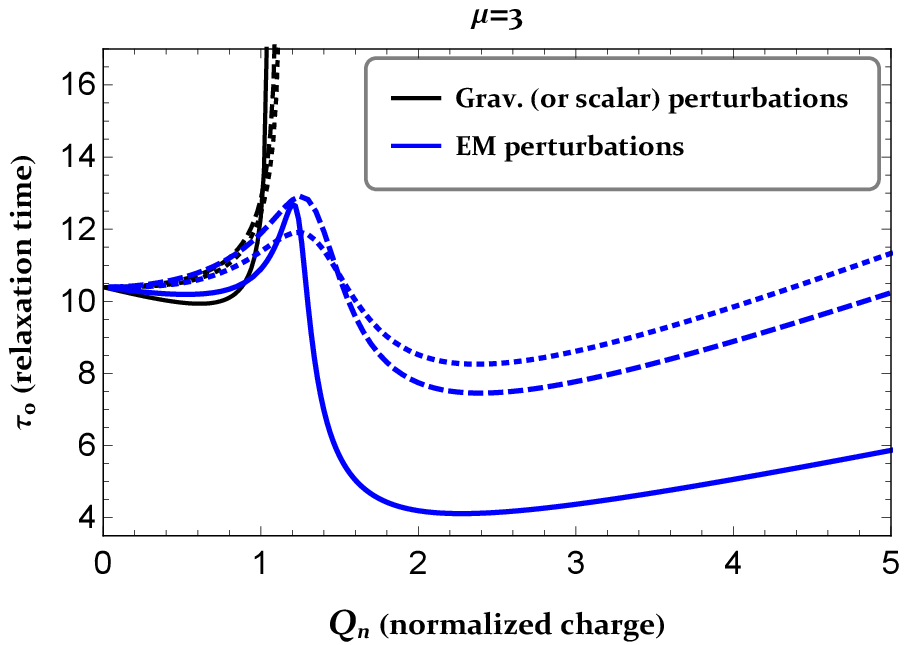}
\caption{\label{fig-r-time} Relaxation times of the gravitational (and scalar) (black) and EM (blue, thick) perturbations of generic class of spacetimes~(\ref{line-element}) with metric function~(\ref{metric-function}) in general relativity coupled to the NED on normalized charge parameters in large multipole numbers limit. Left panel: the Maxwellian regular spacetimes ($\nu=1$) with different values of $\mu$ as $\mu=3$ -- solid, $\mu=5$ -- dashed, $\mu=12$ -- dotted curves. Right panel: the Maxwellian regular spacetimes ($\mu=3$, $\nu=1$) -- solid, the Bardeen regular spacetimes ($\mu=3$, $\nu=2$) -- dashed, the Hayward regular spacetimes ($\mu=3$, $\nu=3$) -- dotted curves.}
\end{figure*}

Thus, from the form of the potential~(\ref{pot-grav1}) one can deduce that in the eikonal regime the scalar and gravitational perturbations of BHs and no-horizon spacetimes in general relativity coupled to NED behave similarly and propagate along null geodesics. Their oscillations and damping rates (or relaxation time) are characterized by the unstable circular null geodesics. At the same time, as it has already been pointed out, the EM perturbations follow the trajectory of light rays or null geodesics of the effective metric~(\ref{eff-metric}). Since in the Maxwellian, Bardeen and Hayward spacetimes of GR with NED there are always nonvanishing circular photon orbits irrespective of the value of the mass and charge parameters, in the eikonal regime BHs and no-horizon spacetimes always oscillate EM perturbations with QN frequencies -- see Figs.~\ref{fig-real} and~\ref{fig-r-time}.

Since the Maxwellian, Bardeen, and Hayward solutions of~(\ref{lagrangian}) reduce to the Schwarzschild one when the charge parameter is set to zero, as $f_{NED}(Q\rightarrow0)=f_{Schw}\equiv1-2M/r$, all the further calculations performed on these spacetimes must coincide with the one that corresponds to the Schwarzschild spacetime in that limit, i.e.
\bear\label{q0limit-correct}
A_{NED}(Q_n\rightarrow0)=A_{Schw},
\ear
where $A$ is any physical quantity. However, at $Q_n=0$ the angular velocities of the circular photon orbits in the Bardeen and Hayward spacetimes are different and they do not match the ones of the Maxwellian BH which coincides with the one of the Schwarzschild BH -- see the right panel of Fig.~\ref{fig-real}, as they have the following limits:
\bear\label{q0limit}
\Omega_{NED}(Q_n\rightarrow0)=\sqrt{\frac{\nu+1}{2}}\Omega_{Schw},
\ear
By comparing (\ref{q0limit-correct}) and (\ref{q0limit}) one easily realizes that only the Maxwellian spacetimes $\nu=1$ have the correct Schwarzschild limit. This confirms once more the fact  that the Bardeen and Hayward solutions have wrong limit at the weak field limit~\cite{Bronnikov:PRL:2000,Burinskii:PRD:2002}. It was shown in our previous papers~\cite{TSSA:PRD:2018,TSA:PRD:98b} that from the imaginary part of the eikonal QNMs of the EM perturbations of BHs in NED one can verify that light rays does not follow null geodesics of the spacetime. Here we have shown that from the real part of the eikonal QNMs of the EM perturbations of spacetimes in NED one can verify if the solution (or equivalently the NED model) has a correct behaviour in the weak EM field limit.

Let us analyze the eikonal QNMs of the Maxwellian, Bardeen and Hayward spacetimes which are presented in Figs.~\ref{fig-real} and~\ref{fig-r-time}. In the eikonal regime the  spacetimes~(\ref{metric-function}) always oscillate the gravitational (scalar) perturbations with bigger real frequency of QNMs than the Schwarzschild one ($\omega_r(Q_n\neq0)>\omega_{r}(Q_n=0)$). The Maxwellian spacetime is the most favourite to oscillate gravitational (scalar) perturbation with bigger real frequency rather than the Hayward spacetime, while the Bardeen one is the least favoured. Moreover, the Maxwellian spacetimes with smaller $\mu$ ($\mu\geq3$) are always better oscillators than the ones with bigger $\mu$.

We consider now the relaxation times of the perturbations in the eikonal regime. In Fig.~\ref{fig-r-time} the relaxation times of the fundamental (the least damped) mode of the gravitational (and scalar) and EM perturbations are presented. In the left panel, the Maxwellian spacetimes with different $\mu$ is shown, while in the right panel the Maxwellian, Bardeen and Hayward spacetimes with $\mu=3$ have been plotted. One can see from the figures that the relaxation times of perturbations of the Maxwellian spacetimes do not depend strongly on the parameter $\mu$. Moreover, the relaxation times of the gravitational (and scalar) perturbations of the regular Maxwellian, Bardeen and Hayward BHs are laid in the similar intermediate ranges, while the ones of the no-horizon spacetimes diverge to infinity at the values which correspond to extreme values of the circular null geodesics. On the other hand the relaxation times of the EM perturbations of these spacetimes qualitatively behave similarly, but quantitatively their differences are significant in the no-horizon spacetimes, i.e., the Hayward no-horizon spacetime oscillates the EM perturbations with the least damping, while the Maxwellian no-horizon spacetime has the fastest relaxation rate.

\begin{table*}
\begin{ruledtabular}
\begin{tabular}{ccccccccc}
Spacetimes & Shortest Grav. & CC & Longest Grav. & CC & Shortest EM & CC & Longest (local) EM & CC\\
\hline
 Schwarzschild BH & 51.0262 & 0 & 51.0262 & 0 & 51.0262 & 0 & 51.0262 & 0 \\
 RN BH$^\ast$ & 50.0378 & 0.73 & 55.5503 & 1 & 50.0378 & 0.73 & 55.5503 & 1 \\
 RN naked singularity & $\gtrsim$55.5503 & $\gtrsim$1 & $\infty$ & 1.06 & $\gtrsim$55.5503 & $\gtrsim$1 & $\infty$ & 1.06 \\
 Maxwellian BH & 48.7728 & 0.62 & 57.5190 & 1 & 50.0385 & 0.52 & 53.4715 & 1 \\
 Maxwellian no-horizon & $\gtrsim$57.5190 & $\gtrsim$1 & $\infty$ & 1.07 & 20.1975 & 2.25 & 62.3036 & 1.2\\
 Bardeen BH & $\gtrsim$51.0262 & $\gtrsim$0 & 63.0313 & 1 & $\gtrsim$51.0262 & $\gtrsim$0 & 58.3588 & 1 \\
 Bardeen no-horizon & $\gtrsim$63.0313 & $\gtrsim$1 & $\infty$ & 1.11 & 36.6051 & 2.4 & 63.1749 & 1.24\\
 Hayward BH & $\gtrsim$51.0262 & $\gtrsim$0 & 61.0489 & 1 & $\gtrsim$51.0262 & $\gtrsim$0 & 55.8758 & 1 \\
 Hayward no-horizon & $\gtrsim$61.0489 & $\gtrsim$1 & $\infty$ & 1.15 & 40.5379 & 2.35 & 58.4751 & 1.24\\
\end{tabular}
\end{ruledtabular}
\caption{\label{tab1} Relaxation times of the gravitational (and scalar) and EM perturbations of the regular spacetimes in comparison with the ones of the RN and Schwarzschild spacetimes in units of $\left[\left(M/M_\odot\right)\mu sec\right]$ microseconds. Where \textbf{CC} stands for the \textbf{C}orresponding \textbf{c}harge. Note that $^\ast$ indicates that in the paper~\cite{Hod:EPJC:2018} by Hod it was shown that in the eikonal regime the RN BH with $Q_n\approx0.73$ has the fastest relaxation rate.}
\end{table*}

In Table~\ref{tab1} we have presented the above discussed features of the relaxation times of the perturbations of the regular spacetimes in unit of $seconds$ in comparison with the ones of the Schwarzschild and RN spacetimes. To write the dimensionful relaxation time in Tab.~\ref{tab1} from dimensionless one in Fig.~\ref{fig-r-time}, one uses the following relation:
\bear
\tau_{ful}=\frac{GM}{c^3}\tau_{less}\approx 4.92\times10^{-6}\tau_{less} \left(\frac{M}{M_\odot}\right) {\rm sec}.
\ear

Since the RN spacetimes is a solution of general relativity coupled with LED, the EM perturbations follow null geodesics in the same manner as the scalar and gravitational ones. Therefore, in Tab.~\ref{tab1} relaxations times of the scalar and gravitational perturbations are identical.

The relaxation times of the nonfundamental modes are easily determined from the relation~\cite{ToshmatovPRD:96}
\bear
\tau_n=\frac{\tau_0}{2n+1}\ .
\ear

\section{Conclusion}\label{sec4}

In this paper we have studied scalar, electromagnetic  and gravitational perturbations of spacetimes in general relativity coupled to the NED. Specifically, we have chosen a  generic model of NED from which the Maxwellian (i.e. corresponding to the Maxwell field in weak field limit), Bardeen and Hayward solutions can be obtained as special cases. In NED light rays do not follow null geodesics of the given spacetime, instead they follow null geodesics of the optical metric. We have shown for the first time that in the Maxwellian, Bardeen and Hayward spacetimes there is always at least one nonvanishing radius for the circular photon orbit around a central gravitating object, while the existence of the circular null geodesics of the spacetime is restricted by the spacetime parameters. These play a fundamental role in the propagation and relaxation periods of the scalar, EM and gravitational perturbations in the eikonal regime. To be more precise, since in the large multipole numbers limit scalar and gravitational perturbations follow the null geodesics of the spacetimes, they behave similarly, therefore, they are indistinguishable from the characteristic frequencies of the perturbations. On the other hand, the EM perturbations follow the light ray trajectory and due to the fact that in the Maxwellian, Bardeen and Hayward spacetimes there is always a nonvanishing circular photon orbit, even no-horizon spacetimes always oscillate the EM perturbations with QNMs.

We have shown that the EM perturbations in the eikonal regime can be a powerful tool to confirm that the light rays do not follow the null geodesics in NED by the relaxation rates and if the underlying solution has a correct weak field limit to the Maxwell electrodynamics (LED) by the angular velocity of the circular photon orbit.

We have shown that the relaxation times of gravitational (and scalar) and EM perturbations of the regular Maxwellian, Bardeen and Hayward BHs are very similar. However, in the horizonless case they behave differently. Interestingly, the RN naked singularity and regular no-horizon spacetimes with the extreme circular null geodesics oscillate the gravitational (and scalar) perturbations with normal modes without damping, i.e., the scalar and gravitational perturbations of these spacetimes never come back to equilibrium. However, the EM perturbations always have damping and they come to relaxation faster than the gravitational ones. Moreover, the relaxation times of the EM perturbations of these spacetimes show qualitatively similar behavior, but quantitatively their differences become significant in the horizonless spacetimes. In other words, the Hayward no-horizon spacetime oscillates the EM perturbations with the least damping, while the Maxwellian no-horizon spacetime is the most favourite in terms of fastness of the relaxation rate.

\section*{Acknowledgments}

This research was supported by the following Grants: Czech Science Foundation GA\v{C}R project No.~19-03950S; Nazarbayev University Faculty Development Competitive Research Grants: ``Quantum gravity from outer space and the search for new extreme astrophysical phenomena", Grant No.~090118FD5348 and by the Ministry of Education of Kazakhstan's target program: ``Center of Excellence for Fundamental and Applied Physics", IRN:~BR05236494; Uzbekistan Ministry for Innovation Development Grants No.~VA-FA-F-2-008 and No.~YFA-Ftech-2018-8, Abdus Salam ICTP through Grant No.~OEA-NT-01 and Erasmus+ exchange grant between Silesian University in Opava and National University of Uzbekistan.

\label{lastpage}
\bibliography{Toshmatov_references}

\end{document}